\documentclass[structabstract]{aa} 
\usepackage{lineno}
\usepackage{graphicx}
\usepackage{natbib}
\bibpunct{(}{)}{;}{a}{}{,} 

\begin{document}
   \title{A Dynamical N-body Model for the Central Region of $\omega$ Centauri}

   \author{B. Jalali \inst{1} 
   \and H. Baumgardt \inst{2,3} 
   \and M. Kissler-Patig \inst{1} 
   \and K. Gebhardt \inst{4} 
   \and E. Noyola \inst{5,6}   
   \and N. L\"{u}tzgendorf \inst{1}
   \and P. T. de Zeeuw \inst{1,7}
          }

   \institute{European Southern Observatory (ESO), Karl-Schwarzschild-Strasse 2, 85748 Garching, Germany
   \and
              Argelander Institute f\"ur Astronomie (AIfA), Auf dem H\"ugel 71, 53121 Bonn, Germany
   \and              
              University of Queensland, School of Mathematics and Physics, Brisbane, QLD 4072, Australia 
   \and           
              Astronomy Department, University of Texas at Austin, Austin, TX 78712, USA
   \and           
              University Observatory, Ludwig Maximilians University, Munich, D-81679, Germany     
   \and
              Instituto de Astronomia, Universidad Nacional Autonoma de Mexico (UNAM), A.P. 70-264, 04510 Mexico 
   \and
              Sterrewacht Leiden, Leiden University, Postbus 9513, 2300 RA Leiden, The Netherlands}

   \date{Received March 18, 2011; accepted November 7, 2011}

  \abstract
   {Supermassive black holes (SMBHs) are fundamental keys to understand the formation and evolution of their host galaxies. However, the formation and growth of SMBHs are not yet well understood. One of the proposed formation scenarios is the growth of SMBHs from seed intermediate-mass black holes (IMBHs, $10^2$ to $10^5 M_{\odot}$) formed in star clusters. In this context, and also with respect to the low mass end of the $M_\bullet-\sigma$ relation for galaxies, globular clusters are in a mass range that make them ideal systems to look for IMBHs. Among Galactic star clusters, the massive cluster $\omega$ Centauri is a special target due to its central high velocity dispersion and also its multiple stellar populations.}
   {We study the central structure and dynamics of the star cluster $\omega$ Centauri to examine whether an IMBH is necessary to explain the observed velocity dispersion and surface brightness profiles.}
   {We perform direct N-body simulations on GPU and GRAPE special purpose computers to follow the dynamical evolution of $\omega$ Centauri. The simulations are compared to the most recent data-sets in order to explain the present-day conditions of the cluster and to constrain the initial conditions leading to the observed profiles.}
   {We find that starting from isotropic spherical multi-mass King models and within our canonical assumptions, a model with a central IMBH mass of 2\% of the cluster stellar mass, i.e. a $5.\times10^4 M_{\odot}$ IMBH, provides a satisfactory fit to both the observed shallow cusp in surface brightness and the continuous rise towards the center of the radial velocity dispersion profile. In our isotropic spherical models, the predicted proper motion dispersion for the best-fit model is the same as the radial velocity dispersion one.}
   {We conclude that with the presence of a central IMBH in our models, we reproduce consistently the rise in the radial velocity dispersion. Furthermore, we always end up with a shallow cusp in the projected surface brightness of our model clusters containing an IMBH. In addition, we find that the M/L ratio seems to be constant in the central region, and starts to rise slightly from the core radius outwards for all models independent of the presence of a black hole. Considering our initial parameter space, it is not possible to explain the observations without a central IMBH for $\omega$ Centauri. To further strengthen the presence of an IMBH as a unique explanation of the observed light and kinematics more detailed analysis such as investigating the contribution of primordial binaries and different anisotropy profiles should be studied.}

   \keywords{Black Holes -- globular clusters: individual (Omega Centauri) -- stellar dynamics -- methods: N-body simulations}

   \maketitle
%

\section{Introduction}
There is no doubt about the existence of supermassive black holes (SMBHs) at the center of most galaxies. However, the formation and growth of SMBHs is poorly understood. One of the proposed scenarios is the growth of SMBHs from seed intermediate-mass black holes (IMBHs, $10^2$ to $10^5 M_{\odot}$) \citep{2001ApJ...562L..19E,2009ApJ...696.1798T}. IMBHs in star clusters might form through the runaway merging of massive stars \citep{2002ApJ...576..899P,2004Natur.428..724P}.

 IMBH formation in star clusters could help to explain the supermassive black hole formation and growth in the center
of galaxies. For instance, \citet{2006ApJ...641..319P} simulate the inner 100 pc of the Milky Way to study the formation and evolution of the population of star clusters and IMBHs in the bulge. They find that 10\% of the clusters born within 100 pc of the Galactic center undergo core collapse during their inward migration and form IMBHs via runaway stellar merging. The IMBHs continue their inward drift towards the Galactic center after the dissolution of the host clusters. \citet{2006ApJ...641..319P} predict that a region within 10 pc of the Galactic center might be populated by 50 IMBHs of about 1000~$M_{\odot}$ mass. They also predict that there is a steady population of several IMBHs  within a few milliparsecs of the Galactic center. This population merges with a rate of about one per 10 Myr with the central SMBH, which is sufficient to build the accumulated majority of the SMBH mass. In the same context, nuclear star clusters co-exist with massive black holes \citep{2008ApJ...678..116S,2010ApJ...714..713S}. The star cluster $\omega$ Centauri (NGC~5139) in our galaxy might be a bridge between smaller systems such as classical globular clusters and larger systems like nuclear star clusters.

If IMBHs form in large numbers in star clusters, then one might expect that some star clusters in the Milky Way or other nearby galaxies contain central black holes.  For example, \citet{2011A&A...533A..36L} report a kinematic evidence for the exsitence of an IMBH in NGC 6388. The structural parameters of globular clusters harboring IMBHs are also studied in \citet{2004ApJ...613.1133B,2004ApJ...613.1143B,2005ApJ...620..238B} and \citet{2010ApJ...720L.179V}. Care should be taken when interpreting cluster morphological parameters as IMBH indicators. For instance, \citet{2007MNRAS.379...93H} shows that the large cores observed in some Galactic star clusters can be caused by heavy stellar mass black hole binaries without the need to invoke an IMBH.
Baumgardt et al. (2005) show that core-collapsed globular clusters with steep surface brightness profiles are not good candidates for harboring central black holes. They find that a cluster hosting an IMBH appears to have a relatively large core with a projected surface brightness only slightly rising toward the center.  It should be noted that Vesperini \& Trenti (2010) argue that shallow cusps in the central surface brightness profile may not be a unique IMBH indicator (see also \citet{2011arXiv1108.4425N} for a different interpretation). Baumgardt et al. (2005) show that the velocity dispersion of the visible stars in a globular cluster with a central black hole remains nearly constant well inside the apparent core radius. Further, they report that in a cluster containing an IMBH, the influence of the black hole becomes significant only at a fraction  $\frac{5}{2}\cdot\frac{M_{BH}}{M_{C}}$ of the half-mass radius (where $M_{BH}$ and $M_{C}$ are the mass of the IMBH and the cluster), i.e. deep within the core, which will affect only a small number of stars.

The star cluster $\omega$ Centauri (NGC~5139), with an estimated mass of $2.5\times10^6 M_\odot$ \citep[][hereafter vdV06]{vdV06} and a tidal radius of about 70 pc (Harris 1996), is the most massive and one of the most spatially extended Galactic star clusters. It has one of the highest central velocity dispersions among the Milky Way star clusters with about $22$ km/s \citep[][hereafter N10]{1995A&A...303..761M, 2010ApJ...719L..60N}. Furthermore, vdV06 measure a rotation of 8 km/s at a radius of about 11 pc from the center using radial velocities. In addition, $\omega$ Centauri is one of the first Galactic globular clusters that have multiple stellar populations among both red giant branch stars \citep{1975ApJ...201L..71F} and main sequence stars \citep{2002ASPC..265...87A, 2004ApJ...605L.125B}. The nature of this cluster is therefore a matter of debate, it could either be a giant globular cluster or the core of a stripped dwarf galaxy \citep{1993ASPC...48..608F,2002ASPC..265....3M,2006ApJ...637L.109B}. The above spectacular properties, in addition to a shallow cusp in the surface brightness profile \citep[][hereafter NGB08]{2008ApJ...676.1008N} and a central sharp rise in the radial velocity dispersion (N10), make $\omega$ Centauri an interesting candidate for harboring a black hole.

$\omega$ Centauri's dynamics is among the best studied of any Galactic star cluster. vdV06 determine its dynamical distance, inclination, mass-to-light ratio, and the intrinsic orbital structure by fitting axisymmetric dynamical models to the ground-based proper motions of van Leeuwen et al. (2000) and line-of-sight velocities from independent data-sets. They find that $\omega$ Centauri shows no significant radial dependence of M/L, consistent with its relatively long relaxation time. Their best-fit dynamical model has a stellar V-band M/L of $2.5 \pm 0.1$ (solar units) and an inclination $i=50^{\circ} \pm 4^{\circ}$, which corresponds to an average intrinsic axial ratio of $0.78 \pm 0.03$. These models do not include any kinematical data in the central $10^{\prime\prime}$. 
\citet{2003MNRAS.339..486G} use Monte Carlo simulations to model $\omega$ Centauri with simple spherical models (neglecting rotation and binary stars). They fit  the surface brightness and radial velocity dispersion relatively well, though again, neither the data nor the model have sufficient resolution in the central 1 pc ($\sim 43^{\prime\prime}$ at 4.8 kpc).

There are several well established correlations between the central black hole mass of galaxies and other parameters of host galaxies such as velocity dispersion \citep{2000ApJ...539L...9F,2000ApJ...539L..13G, 2009ApJ...698..198G}, bulge mass and bulge luminosity \citep{1998AJ....115.2285M, 2004ApJ...604L..89H}. If we extrapolate the Magorrian et al. (1998) relation to the globular cluster mass regime, it predicts an IMBH of about $1.5\times10^4 M_{\odot}$ for $\omega$ Centauri assuming a total cluster mass of $2.5\times 10^6 M_{\odot}$. NGB08 find a $(4\pm1)\times10^4 M_{\odot}$ IMBH applying isotropic Jeans models and a $(3\pm1)\times10^4 M_{\odot}$ IMBH using axisymmetric orbit-based models. More recently, N10 provide new central kinematics of $\omega$ Centauri and suggest a $(5.2\pm0.5)\times10^4 M_{\odot}$ IMBH assuming spherical isotropic Jeans models with respect to a newly determined kinematic center. In contrast, \citet[][hereafter vdMA10]{2010ApJ...710.1063V}, using HST proper motions find a $(8.7\pm2.9)\times10^3M_{\odot}$ IMBH assuming cusp models and an upper limit of $7.4\times10^3 M_{\odot}$ at $1\sigma$ confidence assuming core models (flat central density), while isotropic models imply an IMBH mass of $(1.8\pm 0.3)\times10^4M_{\odot}$. One of the main reasons for the discrepancy is the different cluster centers these two groups used. In addition to the center determination, underestimating the rotation, particularly in the central parsec, could have an important effect on velocity dispersion measurements.

In this paper, we compare the most up-to-date observed surface brightness and kinematic profiles of $\omega$ Centauri with direct N-body simulations in the same way as observers do. This means the same luminosity weights and magnitude cut-offs as in the observations are applied to compute the velocity dispersion and surface density profiles from the models. Similar studies have been performed earlier for the globular cluster M15 in the Milky Way and G1 in M31 by \citet[][]{2003ApJ...582L..21B,2003ApJ...589L..25B}. Direct N-body simulations of M15 explain the observations with a concentration of dark remnants, such as massive white dwarfs and neutron stars in the central regions through mass segregation. Therefore, the presence of an IMBH was not necessary in order to explain the observations. The same conclusion is made by \citet{2006ApJ...641..852V} for M15 using Schwarzschild model. In the case of G1, \citet{2003ApJ...589L..25B} reproduce the observations by assuming a merger history for G1. However, \citet{2005ApJ...634.1093G} provide additional support for the presence of a $2\times 10^4 M_{\odot}$ IMBH. The black hole scenario for G1 is also supported by detections of radio and x-ray  sources in the cluster \citep{2007ApJ...661L.151U, 2006ApJ...644L..45P}.

Here, we examine different IMBH masses (including the no black hole case) in our N-body models with the aim of reproducing the observations for $\omega$ Centauri. Only $N$-body models allow realistic inclusion of relaxation and stellar evolution effects and changes in M/L with radius due to mass segregation. The possible disadvantage of N-body models is that one is restricted to a small number of models since they are time consuming to construct. Therefore, we restrict ourselves to isotropic models and only run a three dimensional grid in concentration ($c$), projected half-mass radius ($r_{hp}$) and IMBH mass space for the scope of this paper. Although axisymmetry and anisotropy are important to include in any modeling, according to vdV06 $\omega$ Centauri is close to isotropic and spherical  within the central few core radii.

In Section 2, we describe the data used in this work to compare with the N-body results. The general recipe for our N-body models is discussed in Section 3. In Section 4 we explain our model results, in particular we discuss the profiles of the no-IMBH and IMBH models and compare them with the observations. We draw our conclusions and discuss possible future work in Section 5.   
\section{Observational Data}
\subsection{The Center of $\omega$ Centauri} 
The determination of $\omega$ Centauri's center is crucial in order to understand and model its dynamics. However, the exact location of the center has been controversial due to the large flat core of the cluster (core radius $\sim100^{\prime\prime}$) and the different methods used to estimate its location.

The center of $\omega$ Centauri has been determined by several authors. Recent determinations are done by NGB08 and \citet[][hereafter AvdM10]{2010ApJ...710.1032A}. 
NGB08 determine the center of $\omega$ Centauri using star counts by excluding the faintest stars due to incompleteness. 
 This method can be biased towards bright stars. In an independent study, AvdM10 determine the
center with different methods including star counts and proper motions. In this case, the authors use star lists
corrected for the presence of bright stars assuming a symmetry axis. This measurement can be biased due to the quality of the correction and the location of symmetry axis. Their result differs from the NGB08 position by $\sim12^{\prime\prime}$. AvdM10 also determine the center of $\omega$ Centauri using HST proper motion data, which they report to be in agreement with their star count method within the uncertainties. AvdM10 might have underestimated the rotation contribution in their local filter window in the proper motion measurements. They try to estimate global rotation but were limited in the amount they could detect. Their evaluated center based on proper motion could possibly be offset from the true center due to this effect.

Due to the above discrepancy, N10 argue that using the kinematic center rather than the density center is the better choice as starting point for models.  For our N-body models we use the kinematical center derived in N10.

\subsection{Surface Brightness Data}
\citet{1987A&A...184..144M} and \citet{1995AJ....109..218T} compile surface brightness data for $\omega$ Centauri from different sources in the literature: aperture photometry of the central regions from \citet{1956MNRAS.116..570G} and \citet{1979AJ.....84..505D} and star counts for larger radii from \citet{1968AJ.....73..456K}. The star counts are characterized by a magnitude limit of B=19 mag. We use the star catalog of AvdM10 to perform star counts in the central regions ($R < 20^{\prime\prime}$) with respect to the kinematic center described above. We use stars brighter than 19.5 mag and an adaptive kernel density estimator for the star counts. The magnitude cut is applied to limit the incompleteness. Our profile center is stable for this magnitude cut-off (fainter magnitude cut-offs cause the density center to shift towards the Anderson center). We adjust our star count profile to the Meylan (1987) and Trager (1995) profiles at larger radii, as taken from NGB08. In NGB08 a surface brightness profile was obtained with integrated light from HST-ACS data within the central $40^{\prime\prime}$ with respect to their center (details explained in \citet{2006AJ....132..447N} and \citet{2007AJ....134..912N}). We use their data from radii larger than $20^{\prime\prime}$, i.e. the data from $20^{\prime\prime}$ to $40^{\prime\prime}$ comes from integrated light measurements and the inner $20^{\prime\prime}$ comes from star counts.
\subsection{Kinematical Data}
N10 obtain kinematics in the central region of $\omega$ Centauri using integral field spectroscopy.  They measure the velocity dispersion from integrated light using VLT-FLAMES with a spectral resolution of $\sim $10,000 in the Ca-triplet wavelength. They tile around the two proposed centers by NGB08 and AvdM10 with eight pointings.
We also use Gemini-GMOS data which NGB08 obtain with integrated light using the same approach as for the VLT-FLAMES data. We use the integrated light velocity dispersion with respect to the kinematic center as presented in Table 1 of N10.

\begin{table*}
\begin{minipage}[]{\textwidth}
\caption{Initial parameters, main assumptions and observed properties used in our models.}
\label{table:2}     
\centering
\renewcommand{\footnoterule}{}  
\begin{tabular}{c c c}        
\hline\hline                 
Property & Symbol & Values\\    
\hline                      
   Num. of stars & N\footnote{see Section 3 for scaling description.} & $5\times10^4$\\   
   Structure Model & -- & King (1966)\\
   Initial concentration & $c$ & 0.3 - 0.8\\
   Initial half-light radius & initial $r_{hp}$ & 11.6 - 14.0 pc\\
   Initial mass function & Kroupa 2001 & 0.1 - 100 $M_{\odot}$\\
   Tidal field & -- & none\\
   Primordial binaries & -- & none\\
   Primordial mass segregation&  -- & none\\
   Mean metallicity& $[Fe/H]$ & -1.62 (Harris 1996)\\

\hline
\end{tabular}
\end{minipage}
\end{table*}
vdV06 collect individual velocity measurements at larger radii from four different sources (\citet{1996AJ....111.1913S}; \citet{1997AJ....114.1087M}; \citet{2006A&A...445..503R}; Xie and Gebhardt (private communication)). Almost all of the above authors measure the velocities of luminous (giants) stars. vdV06 perform many tests such as cluster membership, excluding velocities with large uncertainties and also corrections for perspective rotation, in order to pick only suitable velocities for dynamical modeling. They bin the measurements and obtain the velocity moments in a set of apertures in the plane of the sky. We use the velocity dispersions presented in their Table 4 for comparison with our simulations.

Proper motions are very useful to better constrain the internal dynamics of star clusters, in particular the degree of anisotropy. In addition to ground-based data from vdV06, HST proper motions are available from AvdM10. These authors use isolated stars brighter than the apparent magnitude 21 in their high quality sample for proper motions. In total, they have about 72,000 stars in two fields: one on the cluster center and one positioned adjacent to the first field along the major axis. The central field covers the central $147^{\prime\prime}$ in radius, and the major axis field covers radii between about $100^{\prime\prime}$ to $347^{\prime\prime}$. They use 25,167 stars at $R < 71^{\prime\prime}.7$,  aiming at having the complete position angle coverage in order to calculate average kinematical quantities over circular annuli. However, they stress that the whole data set is usable but is excluded in their main study because of sparse position angle coverage at larger radii.

To compare with the N-body models in this work, we use the proper motions on the minor and major axes available in Table 4 of AvdM10 transformed with respect to the kinematic center in N10. We measure the proper motion dispersion along each axis using a maximum likelihood technique in radial bins taking uncertainties into account \citep{1993ASPC...50..357P}.

Throughout this work, we assume a heliocentric distance of $4.8\pm0.3$ kpc for $\omega$ Centauri (vdV06). Therefore, 1 pc corresponds to $42.97^{\prime\prime}$ in our simulations.
\section{$N$-body Modeling Method}
We started running simulations on GRAPE special purpose computers at ESO using the NBODY4 code \citep{1999PASP..111.1333A} in order to model the star cluster $\omega$ Centauri. It became possible in the middle of this project, however, to take advantage of the recently installed GPU cluster of the University of Queensland which speeds up the simulations by a factor of about 10, allowing us to probe a larger initial parameter space considerably cheaper in computational time. All the results of our models for $\omega$ Centauri are based on simulations using the NBODY6 code \citep{2003gnbs.book.....A} on the GPU cluster.

We follow the method described in \citet{2003ApJ...582L..21B,2003ApJ...589L..25B,2005ApJ...620..238B} to model the dynamical evolution of $\omega$ Centauri. We set up our model clusters following a spherical isotropic King model \citep{1966AJ.....71...64K} in virial equilibrium. The initial stellar masses are drawn from a \citet{2001MNRAS.322..231K} initial mass function (IMF) with lower and upper mass limits of 0.1 and 100 $M_{\odot}$, respectively. Such a choice of IMF is supported in Section 4 since it reproduces well the M/L profile (see Section 4.2), consistent with other independent studies (vdV06). Primordial binaries are not included in our simulations.
 This is justified since $\omega$ Centauri has a very high velocity dispersion, which reduces the contribution of a reasonable small fraction of primordial binaries. In addition,  including primordial binaries is computationally very time consuming and we will investigate it in a separate project. We also do not include initial mass segregation. Furthermore, we do not consider the tidal field of the Galaxy since we are interested in the very central part of the cluster. Neglecting the influence of the tidal field, the evolution of our modeled clusters is driven mainly by two-body relaxation and stellar evolution. We note that mass loss  has little effect on the current velocity dispersion profile if the mass loss occurred early in the evolution. In our simulations, stellar evolution is modeled according to \citet{2000MNRAS.315..543H}. We assume $ [Fe/H]=-1.62 $ dex as the mean cluster metallicity \citep{1996AJ....112.1487H}. The assumed neutron star and black hole retention fraction is set to $10\%$ for both no-IMBH models and models with an IMBH. The initial parameters in our simulations are summarized in Table~\ref{table:2}.

$\omega$ Centauri has a mass of $\sim2.5\times10^6 M_{\odot}$ and therefore $\sim5\times10^6$ stars. Since direct N-body simulations can at the moment handle only clusters with up to $10^5$ stars \citep[see e.g.][]{2010arXiv1010.2210H}, we perform simulations of smaller-N clusters (but more extended in size) and scale our results up to $\omega$ Centauri such as to have the same relaxation time and size as our observed cluster after 12 Gyr of evolution.  The relaxation time of a cluster with mass $M$ and half-mass radius $r_h$ is given by \citet{1987degc.book.....S} as:
\begin{equation}
T_{r_{h}} = 0.138 \frac{\sqrt{M} \; r_h^{3/2}}{\langle m \rangle \;\sqrt{G} \; \ln (\gamma N)} \;\; ,
\label{rscale}
\end{equation}
where $\langle m \rangle$ is the mean mass of the stars in the cluster, $N$ is the number of stars, and $\gamma$ is a factor in the Coulomb logarithm, approximately equal to 0.02 for multi-mass clusters \citep{1996MNRAS.279.1037G}.
 The scaling factor for the positions is given by
\begin{equation}
r_{scale} =  \frac{r_{hp ocen}}{r_{hp sim}} \;, 
\end{equation}
where $r_{hp ocen}$ is the projected half-light radius of $\omega$ Centauri \citep[5.83 pc at 4.8 kpc distance,][]{1996AJ....112.1487H} where the integrated cluster light reaches half its maximum value and $r_{hp sim}$ is the projected half-light radius of model clusters. In order to have the same relaxation time as the observed cluster, we have to scale up the mass of our clusters to a mass $M_{ocen}$ which satisfies the following equation:
\begin{equation}
\left( \frac{M_{sim}}{M_{ocen}} \right)^{1/3} \left( \frac{\ln (\gamma M_{ocen})} {\ln (\gamma M_{sim})} \right)^{2/3} = \frac{r_{hp ocen}}{r_{hp sim}} \;.
\end{equation}
Here $M_{ocen}$ is the bound mass of the models at the end of our simulations (T = 12 Gyr). Since the size and mass of the model clusters are re-scaled, we also scale the velocities of the stars by a factor:
\begin{equation}
v_{scale} = \left( \frac{r_{hp sim}}{r_{hp ocen}} \right)^{1/2} \; \left( \frac{M_{ocen}}{M_{sim}} \right)^{1/2} \;, 
\end{equation}
where subscripts ``ocen" and ``sim" denote the actual values for $\omega$ Centauri and those in the simulations, respectively \citep[see][]{2003ApJ...589L..25B}. In eq. (4), the first factor is needed due to the reduction of distances between stars while the second factor takes care of the increase in cluster mass when scaling our models to $\omega$ Centauri. Due to eqs. (3) and (4) models starting with larger initial $r_{hp sim}$ will end with a higher cluster mass and therefore higher velocities (see Section 4).
\begin{figure}[b!] 
   \centering
   \includegraphics[width=0.5\textwidth]{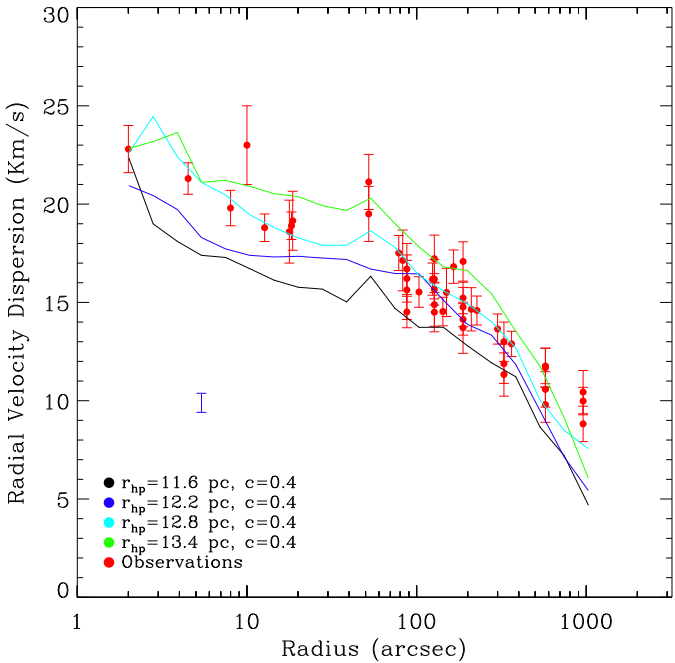}
      \caption{Radial velocity dispersion profiles as a function of radius for models with an IMBH mass of 2\% of the stellar mass with the same initial concentration factor. Red points are the observed data taken from different sources (section 2.3). Black, magenta, light blue and green are models with initial half-light radius ($r_{hp}$) of 11.6, 12.2, 12.8 and 13.4 pc, respectively. A typical uncertainty for models at $5^{\prime\prime}$ is indicated in blue. Higher initial $r_{hp}$ models end up with a higher velocity dispersion across all radii but with roughly the same overall shape because of their higher total cluster mass as described in section 4.}
         \label{fixlogC}
   \end{figure}
   \begin{figure}[b!] 
   \centering
   \includegraphics[width=0.5\textwidth]{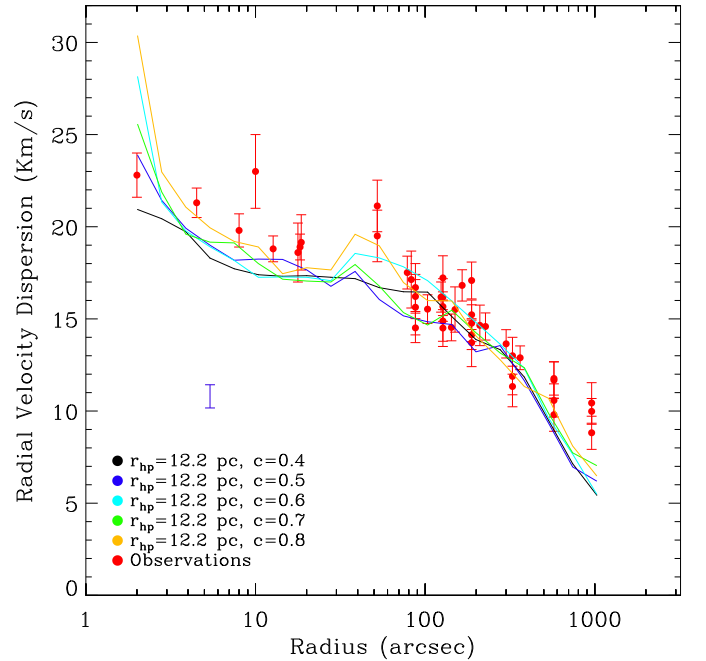}
      \caption{Radial velocity dispersion profiles as a function of radius for models with an IMBH mass 2\% of the stellar mass with the same initial $r_{hp}$. Observational data are shown as red points, and a sample uncertainty is shown in blue as in Figure~\ref{fixlogC}. Black, magenta, light blue, green and orange points are models with initial $c=0.4$ to 0.8. All models have the same dispersion outside the core radius ($\sim100^{\prime\prime}$) while clusters with higher concentration show higher velocity dispersion towards the center, as explained in section 4.}
      \label{fixrhp}
   \end{figure}

We perform all simulations with $5\times10^4$ stars, which before scaling gives us an initial cluster mass of $\sim3.5\times10^4 M_{\odot} $ and a final cluster mass of $\sim2.0 \times 10^4 M_{\odot}$. Usually, the bound stars are about $ 99\% $ of the initial number of stars.

In order to find the initial conditions of $\omega$ Centauri that lead to the present-day observed kinematics and surface brightness profile, we set up clusters with different initial half-mass radii, different concentration parameters, defined as $ \log(\frac{r_{t}}{r_{c}})$ where $r_{t}$ is the tidal radius and $r_{c}$ is the core radius, and different IMBH masses and evolve them with NBODY6 up to an age of 12 Gyr. Then, we estimate how closely each model cluster reproduce the observed profiles using a $\chi^2$ test. We calculate a $\chi^2$ value for the surface brightness, the radial velocity, and the proper motion dispersion profiles of each model to compare with the data using
\begin{equation} 
\chi^2 = \sum_{i = 1}^{N} \left( \frac{{\cal M}_{i}-{\cal D}_{i}}{\sqrt{ (\Delta{\cal M}_{i})^2 + (\Delta{\cal D}_{i})^2 }} \right)^2.
\end{equation}
${\cal M}_{i}$, ${\cal D}_{i}$, $\Delta{\cal M}_{i}$ and $\Delta{\cal D}_{i}$ are the model and data points and their relevant uncertainties. $N$ is the number of data points for the radial velocity dispersion, the proper motion dispersion and the surface brightness profiles. We calculate the absolute $\chi^2$ values for all the $\chi^2$ maps in the next section. 
We use $\chi^2$ values only to quantify judgments on different model profiles in comparison with observed ones. We calculate the $\chi^2$ values for all models within the inner $400^{\prime\prime}$ ($\sim 10$ pc) since we do not consider tidal fields and this in turn affects the number of stars at larger radii. We aim to simultaneously reproduce the observed velocity dispersions and surface brightness profiles. Therefore, we apply the following relation to obtain the reduced combined $\chi^2$ values of radial velocity, and proper motion dispersion and surface brightness for each model:
\begin{equation} 
\chi^2 = \frac{\chi^2_{rv} + \chi^2_{pm} + \chi^2_{sb}}{N_{rv} + N_{pm} + N_{sb}} \;,
\end{equation}
``rv", ``pm" and ``sb" stand for radial velocity dispersion, proper motion dispersion and surface brightness, respectively. 
$N_{rv}$, $N_{pm}$, and $N_{sb}$ are 38, 218, and 32 over $400^{\prime\prime}$. 
 We note that the large difference between the number of ``rv", ``pm" and ``sb" data points will cause a non-smooth total $\chi^2$ space between different models and consequently results in small absolute $\chi^2$ differences, as we see in Section 4.2.

We vary the initial parameters to compare the resulting profiles with observations of $\omega$ Centauri. We vary the IMBH mass between 0\% to 3\% of the model stellar mass. In total we compute more than 100 models to find the initial conditions which reproduce the observations of $\omega$ Centauri best. The final quantities (e.g. number density and velocity) are calculated  by adding up three model clusters with the same physical initial conditions but different random number seeds and also superimposing 10 snapshots in the case of no-IMBH models and 20 snapshots in the case of IMBH models to improve the statistics. The snapshots start at 11 Gyr with a 50 Myr step.
\subsection{Model Density Profile}
For each model we calculate the surface brightness and kinematic profiles including radial velocity and proper motions, using similar magnitude limits as observers use for $\omega$ Centauri. 
In order to determine the surface brightness profile, the density center of our model clusters is determined using the method of \citet{1985ApJ...298...80C}. We then count the number of bound stars in two-dimensional circular annuli around the density center. We use the infinite projection method of \citet{2005ApJ...619..243M} and average each quantity over all (infinite) orientations for each bin based on geometric arguments. Infinite projection gives significantly better statistics over using only a finite number of projections, especially  in the inner cluster parts. We convert the star counts per parsec squared to numbers per arcsecond squared using the assumed distance to $\omega$ Centauri. 
We bin the stars around the (density) center in 20 annuli of equal logarithmic width between $2.0^{\prime\prime}$ and $1000^{\prime\prime}$.

We separately calculate the surface number density of all stars (including dark remnants) and only bright stars.
In order to compare the models with the observations, we only consider stars brighter than V=22 magnitude. We convert model bolometric luminosities to V-band luminosities assuming the stellar temperature model of \citet{1989ApJ...347..998E}. Using the distance modulus of $\omega$ Centauri,  we obtain V-band magnitudes that can be directly compared to the measurements (e.g. magnitude cuts). AvdM10 measure the observed surface brightness profile using HST multi-epoch data with the magnitude cut of B$\sim$22 mag (section 3.2 and their Figure 1). Our magnitude cutoff of V=22 magnitude is also consistent with the combined data in \citet{1987A&A...184..144M} and \citet{1995AJ....109..218T}. The adopted cutoff in our simulations is also applied to maximize the number of stars in each bin for better number statistics. We then convert the star counts to magnitude per arcsecond squared. We match our surface brightness profile with the observed one by shifting it by an additional zero point. We calculate the zero point for each model profile with a $\chi^2 $ minimization in order to compare our profile directly to the observed surface brightness profile.
\subsection{Model Kinematic Profile}
We calculate the velocity dispersion profile of our model clusters again using all stars and using only bright stars similar to how the surface brightness profile was calculated. First, we determine the velocity dispersion using all stars including compact remnants. Second, we use only stars brighter than a certain magnitude limit, adopted to be the same as the observational one.  In the case of radial velocities, the observed kinematical data within the central $30^{\prime\prime}$ are obtained using integrated light (IFU data), so we similarly measure in the models the luminosity weighted velocity dispersions. We avoid very bright stars in the central region as in the observations, in order to minimize shot noise effects.
Therefore, we only consider stars fainter than V-magnitude of 15 (similar to the observational cut off) inside  $30^{\prime\prime}$. For radial velocities at larger radii, we consider stars brighter than an apparent magnitude of V=18 magnitude since the observed data were mainly obtained from individual giant stars.
For proper motions, we consider stars within the magnitude range of $18 < V < 22$ magnitude which is found suitable for proper motion measurements in AvdM10. We assign magnitude weights to the velocities of stars within this magnitude range following Table 4 of AvdM10 to measure the proper motion velocities.
\begin{figure}[b!]
   \centering
   \includegraphics[width=0.5\textwidth]{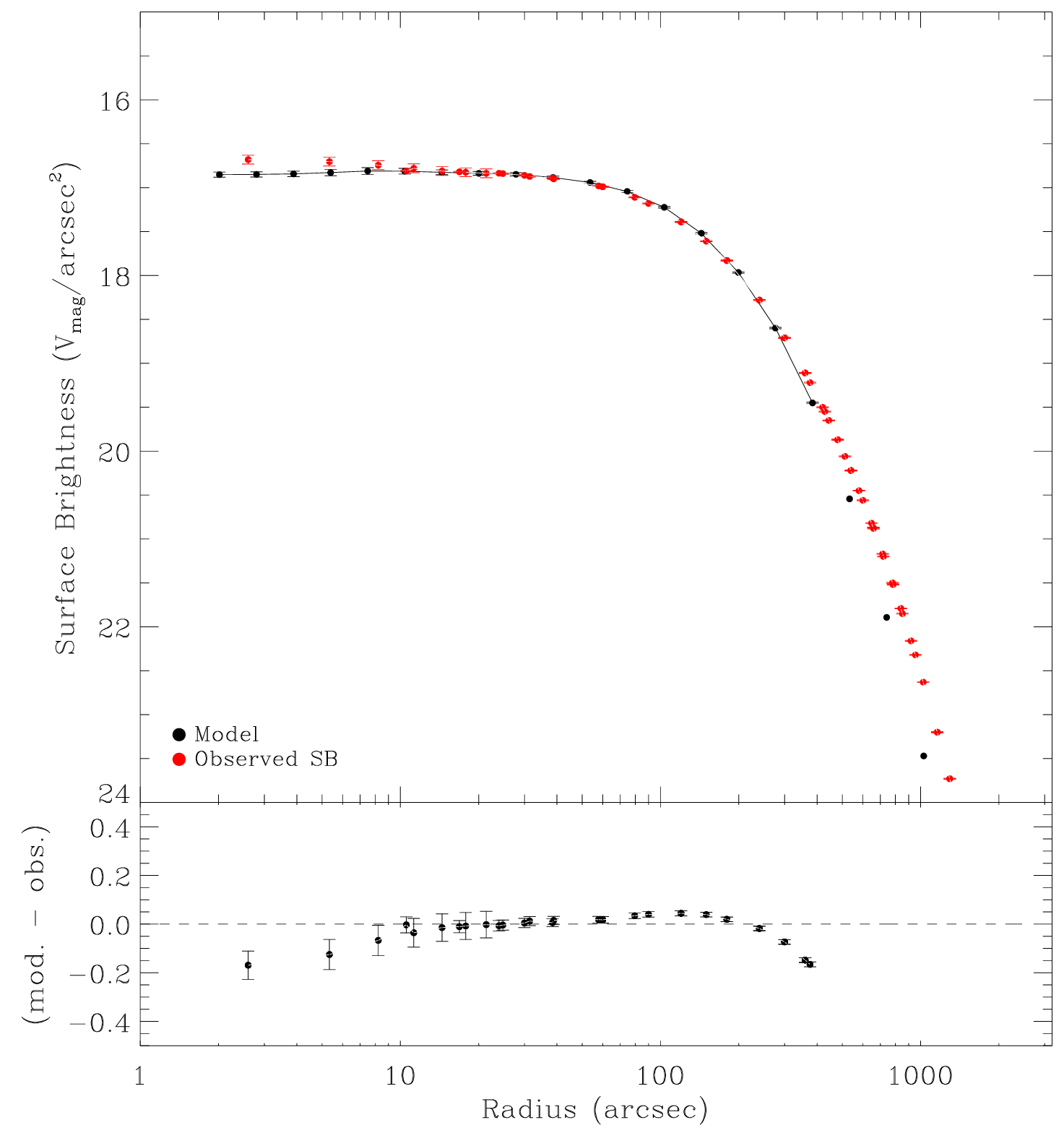}
      \caption{Upper panel: V-band surface brightness profile as a function of radius. The red points are the observed V-band surface brightness relative to the kinematic center in N10. The black points are the best no-IMBH model. This model fits the surface brightness data relatively well inside $400^{\prime\prime}$, except inside the central $10^{\prime\prime}$, where it ends up below the observations. The model points are connected inside $400^{\prime\prime}$ over which we calculate the $\chi^2$ values. Lower panel: residual of our model with respect to the observed profile.}
         \label{SB}
   \end{figure}
\section{Results}
In order to find a model which has simultaneously a good fit of the observed surface brightness and velocity dispersion profiles of $\omega$ Centauri, we run a grid of models with different initial conditions. We describe below some general phenomena in order to illustrate the effect of each initial parameter on the profiles of the evolved clusters. We use as an example the IMBH model with 2\% mass of the total cluster mass. 
We first consider variations of the initial $r_{hp}$ while the cluster concentration $(c)$ is fixed. The model cluster final mass is a free parameter as discussed for eq. (3). Since the final $r_{hp}$ is fixed to the $\omega$ Centauri one (4.18 arcmin in the 2003 version of \citet{1996AJ....112.1487H}), increasing the initial radius will produce a more massive cluster after scaling, since we scale such as to keep the relaxation time constant (see eq. 3). Hence, at a fixed initial cluster concentration, by increasing the initial $r_{hp}$ the whole radial velocity dispersion profile will scale up, as can be seen in Fig~\ref{fixlogC}. In this figure, at a fixed concentration $c=0.4$ one can see that a cluster which starts with an initial $r_{hp}$ of 11.6 pc has a much lower velocity dispersion profile than the observed one at almost all radii. In contrast, the cluster with the same concentration but higher initial $r_{hp}$ of 13.4 pc has a higher dispersion profile than the observed one, while the general shape of the profiles usually follows the same pattern. This also shows that relaxation is not very important for $\omega$ Centauri. 

We now discuss the effect of varying the initial concentration $(c)$ on the final kinematic profiles after 12 Gyr of evolution. Clusters with higher concentrations have more mass in the central regions and therefore a higher central velocity dispersion. In Figure~\ref{fixrhp}, we show this effect by presenting one family of models with fixed initial $r_{hp}$ (12.2 pc) but varying concentrations (see the color code in the caption). In this example,  all clusters in the family of $r_{hp}=12.2$ pc have almost the same velocity dispersion at large radii, but different central velocity dispersions as a function of their concentrations.     

   \begin{figure}[]
   \centering
   \includegraphics[width=0.5\textwidth]{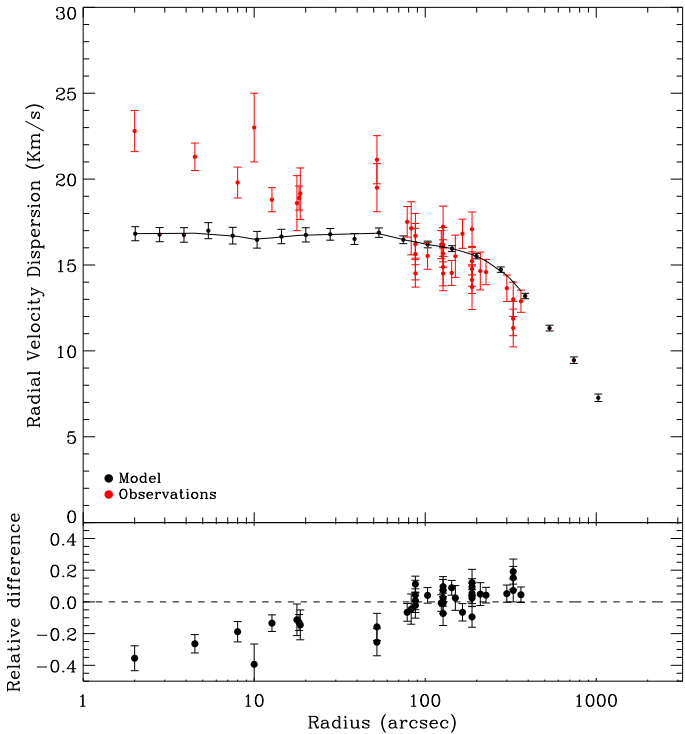}
      \caption{Upper panel: Radial velocity dispersion profile vs. radius. The red points are the observed velocity dispersion relative to the kinematic center, taken from N10. The velocity dispersion of the best-fit no-IMBH model is shown in black. The model obviously does not fit the data around the center (for details see section 4.1). Lower panel: the relative difference of our model and the observed profile.}
         \label{RV}
   \end{figure}
   \begin{figure}[]
   \centering
   \includegraphics[width=0.5\textwidth]{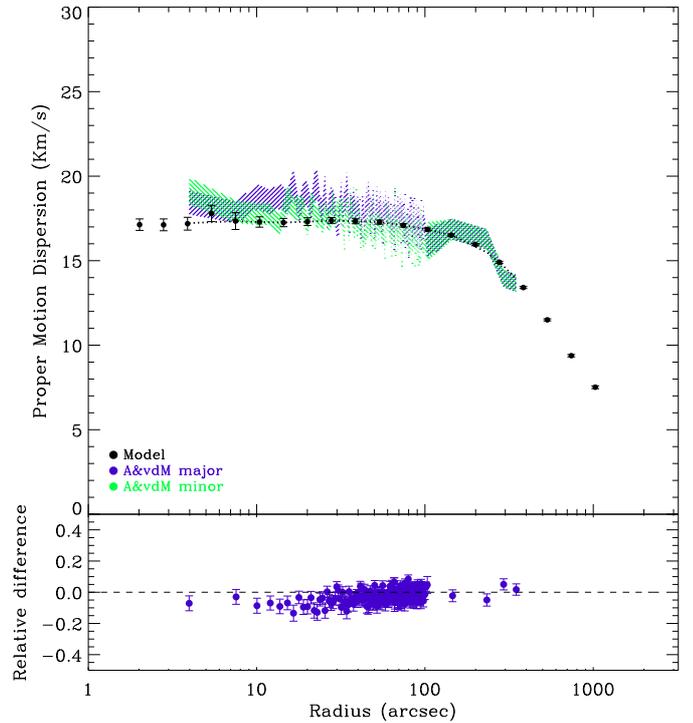}
      \caption{ The proper motion velocity dispersion profile vs. radius. The best-fit no-IMBH model is shown in black. Shaded magenta and green are the observed proper motions for major and minor axes taken from vdMA10 but with respect to the kinematic center. The Lower panel shows the residuals and displays the major axis data only, for clarity.}
         \label{PM}
   \end{figure}
\subsection{No-IMBH Models}
We first run a sparse grid of models between $0.5 <c< 1.5$ and $10.0 <r_{hp}< 15.0$ pc in order to identify the best fitting model. We find that models with initial $c\sim0.8$ and $r_{hp}\sim12$ pc give the best fit. We produce a finer grid of models between $0.6 <c< 1.0$ and $12.2 <r_{hp}< 14.0$ pc for the no black hole case. As explained in section 3, we calculate the $\chi^2$ values of velocity dispersion, proper motion dispersion and surface brightness profiles to choose the best-fitting no-IMBH model. We find that the model with initial $c=0.8$ and initial $r_{hp}=12.8$ pc produces the best fit to the data of $\omega$ Centauri. 

Figure~\ref{SB} shows the surface brightness profile of bright stars with V $<$ 22 magnitude for the model with $c=0.8$ and $r_{hp}=12.8$ pc and no central black hole. In this model, the surface brightness of bright stars reproduces the observational data well but predicts slightly lower values in the central $10^{\prime\prime}$ where NGB08 see a shallow cusp, which can be interpreted as an evidence of an IMBH. Fig.~\ref{RV} shows the radial velocity dispersion profile for the above no-IMBH model. Except for the innermost part, the model agrees relatively well with the data within the uncertainties but in the central $100^{\prime\prime}$ the no-IMBH model lies significantly below the observational data.  
The proper motion dispersion profile for the no-IMBH best-fit model is shown in Fig.~\ref{PM}. 

We note that whether the no-IMBH model does a good or a poor fit depends on the adopted center since in vdMA10 there is no shallow cusp in the observed surface brightness profile. Further, the observed proper motion data around the proposed center by AvdM10 would be better fitted by the no-IMBH models although the central $10^{\prime\prime}$ values are still slightly above the best no-IMBH models.
\begin{figure}[t!]
   \centering
   \includegraphics[width=0.5\textwidth]{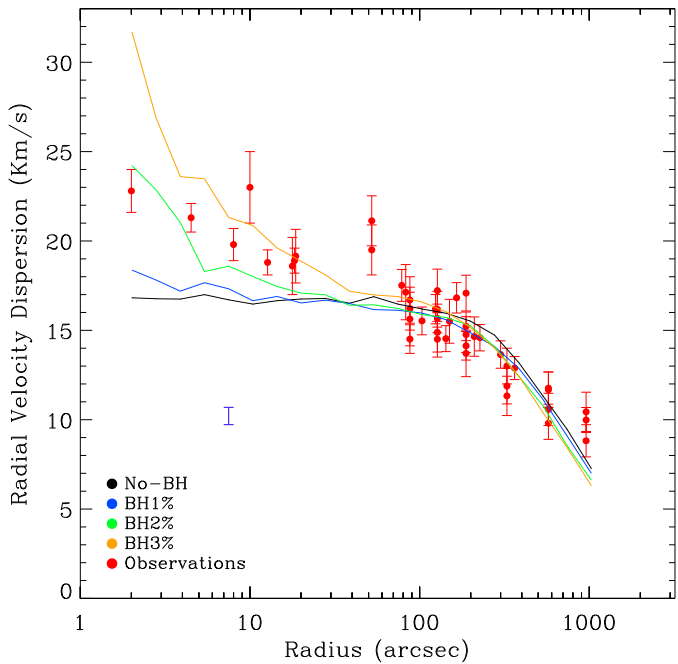}
      \caption{Velocity dispersion profile with different central black hole masses. The red points are the observed dispersion. The black line is the best no-IMBH model which obviously cannot reproduce the data in the inner parts. The blue, green and yellow lines represent models with black hole masses of 1$\%$, 2$\%$ and 3$\%$ of the cluster total mass. A typical uncertainty at $5^{\prime\prime}$ is shown in blue on top of models for visual clarity. The green line is the velocity dispersion profile for the best-fit black hole model, containing 2$\%$ of the initial cluster stellar mass (section 4.2).}
         \label{BHspectra_vel}
   \end{figure}
\begin{figure}[t!]
   \centering
   \includegraphics[width=0.5\textwidth]{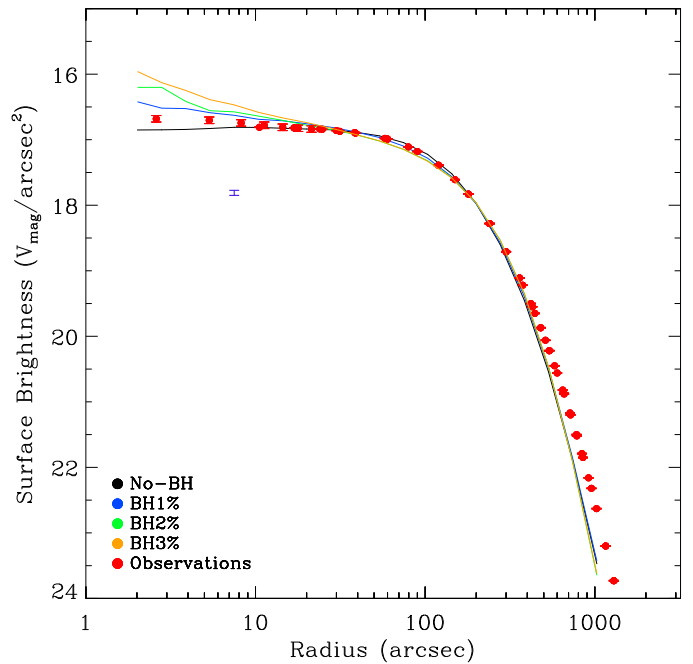}
      \caption{Surface brightness profile of models with different central black hole masses. Symbols and colors are the same as Fig ~\ref{BHspectra_vel}. The red points are the observed profile (section 2). The black line is the best no-IMBH model which falls slightly below the observed data. The blue, green and yellow lines represent models with black hole masses of 1$\%$, 2$\%$ and 3$\%$ of the cluster total mass.}
         \label{BHspectra_SB}
   \end{figure}
   
Models starting with higher $r_{hp}$ than 12.8 pc cannot match the radial (and proper motion) velocity dispersion profile beyond $\sim50^{\prime\prime}$, they always lie higher than the data at larger radii when we attempted to match the inner region. In other words, higher initial $r_{hp}$, meaning higher model cluster mass, increase the whole model profile. Therefore, models with an initial $r_{hp}$ higher than 12.8 pc are ruled out. Similarly, models with 12.2 pc initial $r_{hp}$ do not reproduce the observations either. However, the majority of the data points used to compute the radial velocity dispersion is at radii larger than $\sim100^{\prime\prime}$, while for the proper motion dispersion it is at radii smaller than $\sim100^{\prime\prime}$. This can lead to a slightly different best-fit model depending on the radial range over which the model is evaluated.  Given eq.~(3), we use the data at large radii to choose the best cluster mass scaling. For instance, the model with $c=0.8$ and $r_{hp}=12.8$ pc better fits the proper motions than the $c=0.8$ and $r_{hp}=12.2$ pc which better fits the radial velocity dispersion. The latter provides a good fit to the data, but the total $\chi^2$ is slightly larger than for the former since the number of proper motion data points is much larger (218 points) than the number of data points for radial velocity data (38 points), giving a higher weight to the proper motion fit.
\subsection{IMBH Models}
Since models without an IMBH could not represent the data well, we run models including central black holes of various masses with the hope of improving the fit to the data. In this set of simulations, we start from isotropic King model conditions with concentrations in the range $0.3 < c < 0.8$ and initial $r_{hp}$ in the range of $12.2 < r_{hp} < 14.0$ pc. Apart from the central black hole mass, all the other parameters such as NS retention fraction and stellar mass range are the same as for the no-IMBH models. Furthermore, since we find that the IMBH model that reproduces the observed profiles best has lower concentration and higher $r_{hp}$ than the no-IMBH models, we explore a larger parameter space in order to determine the best set initial conditions leading to the present-day observations. For each initial configuration, we run the simulations as described in Section 3. For models with a central black hole, we calculate the kinematic and surface brightness profiles as described in the Sections 3.1 and 3.2. Hence, all the model-data comparisons are done following the same magnitude and radial cut-offs as for the no-IMBH models. In the case of IMBH models, the IMBH moves in the core of the cluster during the simulations, and we center the final model clusters on the central black hole. In our simulations, IMBH models contain black hole masses of 1$\%$, 2$\%$ and 3$\%$ of the final stellar mass of the cluster. For reference, a 1\% IMBH would lie slightly above the \citet{1998AJ....115.2285M} relation. If we adopt a mass of $\sim~2.5\times10^6M_{\odot}$ for $\omega$ Centauri (vdV06), the IMBH masses we use are equivalent to black hole masses of about $2.5 \times 10^4 M_{\odot}$, $5.0 \times 10^4 M_{\odot}$ and $7.5 \times10^4M_{\odot}$ respectively. However, the exact masses of our models depend on the initial $r_{hp}$ due to the scaling of radii (see eq. 3), and will slightly differ from the above values.
We calculate and analyze the $\chi^2$ maps for models including the above IMBH masses. We find the best-fit models for each grid including a 1\%, 2\% and 3\% IMBH, applying the same methodology that we use to find the best no-IMBH model. The radial velocity dispersion and surface brightness profiles of the best-fit models of different IMBH masses and also the ones for the best-fit no-IMBH model are shown in Figure~\ref{BHspectra_vel} and Figure~\ref{BHspectra_SB}.  
The best-fit IMBH model is the one containing 2\% of the stellar cluster mass starting initially with $c=0.5$ and $r_{hp}=13.4$ pc. As can be seen from Figure~\ref{BHspectra_vel}, the best-fit IMBH models with 1$\%$ and 3$\%$ of the cluster stellar mass do not fit the observed radial velocity dispersion well since they have lower or higher values than the observed velocity dispersion profile in the central part. A summary of best-fit models among all models including no-IMBH and IMBH are tabulated in Table~\ref{table:1}. The chosen best-fit 2\% mass IMBH model is highlighted with boldface.
\begin{figure*}[t!]
    \centering
    \includegraphics[width=\textwidth]{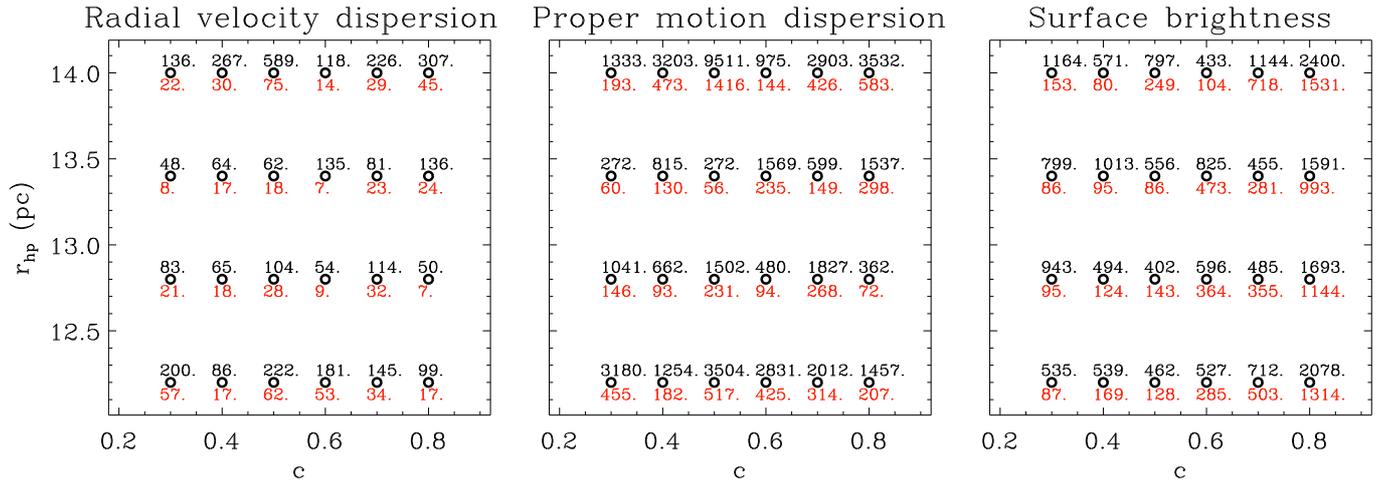}
       \caption{From left to right, the $\chi^2$ map for the radial velocity, the proper motion dispersion and the surface brightness profiles, for models with an IMBH mass of 2\% of the model cluster mass, over the grid of initial parameters space. The numbers in black are the absolute $\chi^2$ over $400^{\prime\prime}$ radius. The absolute $\chi^2$ values inside $40^{\prime\prime}$ are shown in red to visualize the goodness of the fit in the central region for each model.}
          \label{multigrid}
\end{figure*}
\begin{figure}[t!]
   \centering
   \includegraphics[width=0.5\textwidth]{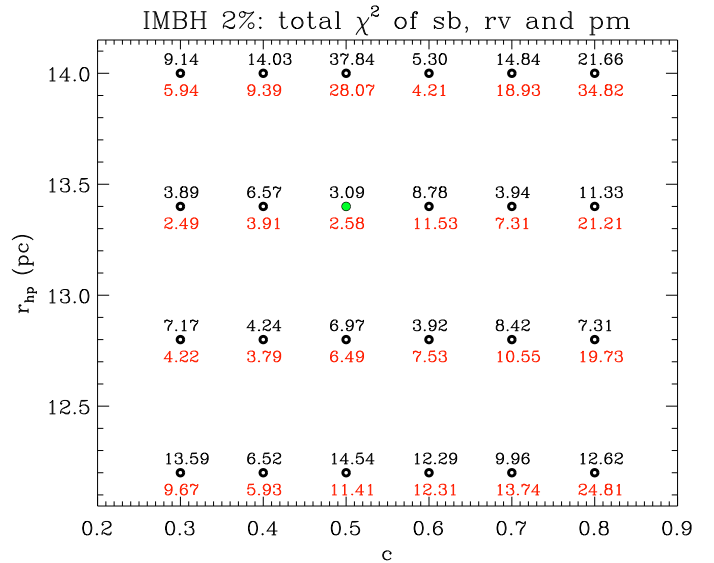}
      \caption{The final reduced, combined $\chi^2$ map for models with an IMBH with a mass of 2\% of the model cluster mass. This map is calculated based on equation 6, using the radial velocity, the proper motion dispersion and the surface brightness $\chi^2$ maps in Fig~\ref{multigrid}. The best-fit model is marked in filled green.}
         \label{finalcombinedchi2}
\end{figure}

The left panel in Figure~\ref{multigrid} shows the radial velocity dispersion $\chi^2$ map for models containing an IMBH mass of 2$\%$ of the cluster stellar mass. Models starting with an initial $r_{hp}$ of 12.2 (14.0) pc always lie lower (higher) than the data at large radii because of the less (more) massive final cluster mass (see eq. 3). In contrast, models starting with $r_{hp}$ of 12.8 and 13.4 fit the data much better at larger radii due to better cluster mass scaling when relaxation time is the same as for $\omega$ Centauri.

\begin{figure}[]
   \centering
   \includegraphics[width=0.5\textwidth]{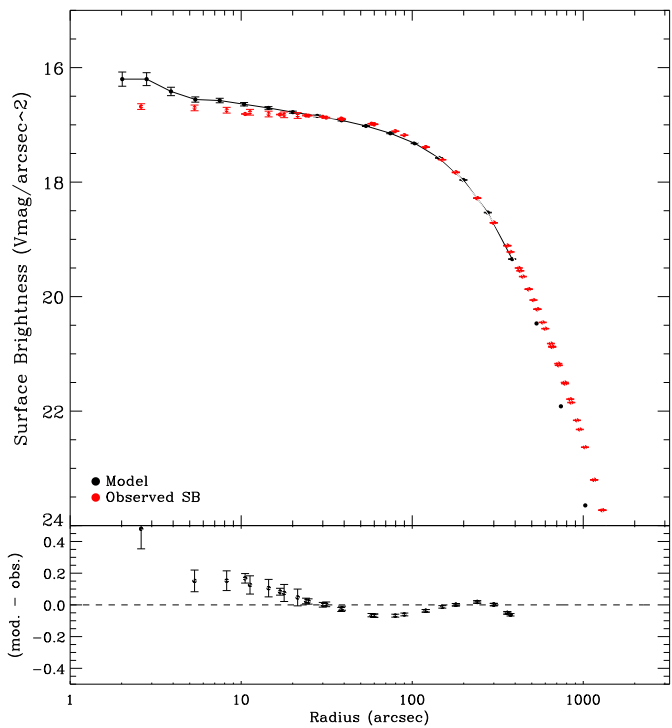}
      \caption{Upper panel: V-band surface brightness profile as a function of radius for the best-fit model with an IMBH mass of 2\% of stellar mass. The symbols are as in Fig.~\ref{Best_BH2RV}. Lower panel: residual of our model and the observed profile.}
         \label{Best_BH2SB}
\end{figure}
In addition to radial velocities, we calculate the proper motion dispersion (perpendicular component to the radial velocity in our models) with the weights and magnitude cut-offs explained in section 2.3 and 3.2.
The $\chi^2$ map for the proper motion dispersion for the 2\% IMBH models is shown in the middle panel of Fig.~\ref{multigrid}.
The right panel in Figure~\ref{multigrid} depicts the $\chi^2$ values of surface brightness for the models with an IMBH mass of 2$\%$ of the stellar mass. Models starting with high initial concentrations such as $c=0.7$ and 0.8 generally show a steeper cusp in the central part than the observation and do not describe the observed surface brightness well, they have higher $\chi^2$ values.

A larger core gives a better fit at larger radii (around $80^{\prime\prime}$), though these models have a very poor fit in the central part due to a very steep rise. In addition, it should be noted that the number of data points at large radii is higher than at the center. Furthermore, the observational uncertainties are smaller at larger radii than in the central region, therefore, the $\chi^2$ value can be small if a model has a good fit at large radii but an unsatisfactory fit in the central part. Thus, we choose models with smaller $(c)$ such as 0.4 and 0.5 as our best-fit models for the surface brightness data. The fact that smaller core models such as $c=0.4$ and 0.5 do not provide a better fit at larger radii in surface brightness is related to the initial King profile, which alone might not be sufficient to fit the density profile and is independent of having a central IMBH. In a future paper we intend to modify the initial density profile so that we take this effect into account, for instance by combining a King profile with a Sersic one.
\begin{figure}[t!] 
   \centering
   \includegraphics[width=0.5\textwidth]{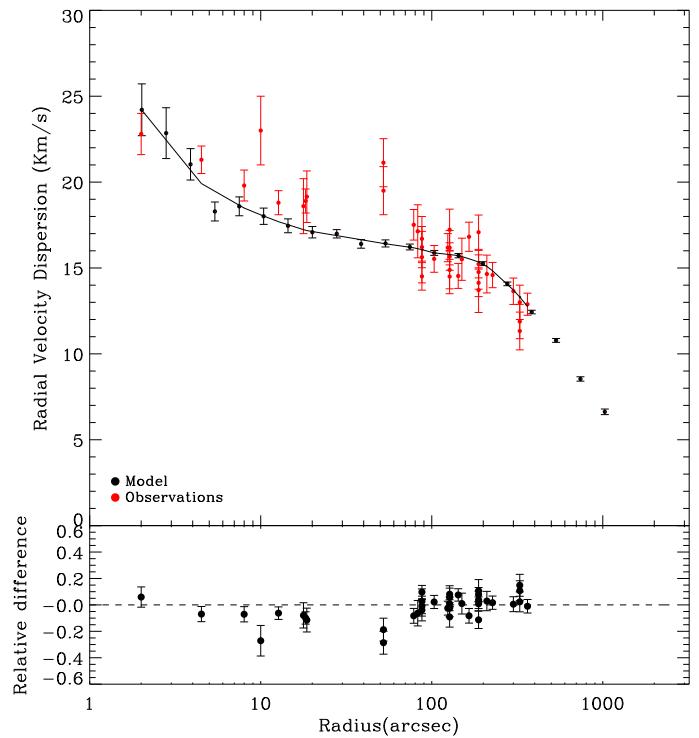}
      \caption{Upper panel: Velocity dispersion profile vs. radius in arcsecond. The red points are the observed data points relative to the kinematic center, taken from N10. The velocities for the best-fit model containing an IMBH mass of 2\% of stellar mass are shown in black. Lower panel: the relative difference between our model and the observed profile.}
         \label{Best_BH2RV}
\end{figure}
\begin{figure}[t!]
   \centering
   \includegraphics[width=0.5\textwidth]{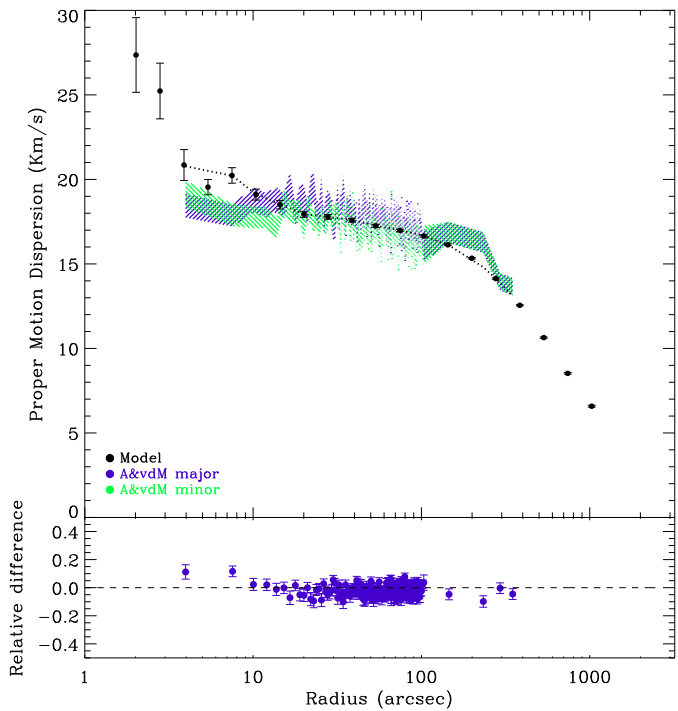}
      \caption{The proper motion dispersion profile for our best-fit IMBH model is shown in black. Shaded magenta and green are the observed proper motions for major and minor axes taken from vdMA10 but with respect to the kinematic center. The model proper motion dispersions for a cluster containing an IMBH mass of 2\% of the stellar mass has a higher velocity dispersions than the observed one in the central $10^{\prime\prime}$. The residual is illustrated only for major axis data for clarity.}
         \label{modelPM_AvdM}
\end{figure}

Considering the above results, the model with initial $c=0.5$ and $r_{hp}=13.4$ pc fits better than the other models. However, it still has a slightly steeper surface brightness in the central region than the observed one. 
We show the final reduced combined $\chi^2$ values (see eq. 6) for models with a 2\% IMBH mass in Figure~\ref{finalcombinedchi2}. The IMBH model starting with initial $c=0.5$ and $r_{hp}=13.4$ pc provides the best-fit to the observations. The profiles for the best-fit IMBH model are shown in Figures~\ref{Best_BH2SB}, ~\ref{Best_BH2RV} and ~\ref{modelPM_AvdM}.
As can be seen in Figure~\ref{modelPM_AvdM}, a model with an IMBH mass of 2\% of the final cluster stellar mass shows higher central velocity dispersions than the observed proper motions. However, as mentioned before, the best-fit no-IMBH model might fit the proper motions better with respect to the AvdM10 center.

Furthermore, it is interesting to note that the radial versus transversal components of the velocities for our evolved clusters are the same. Thus, an initially isotropic cluster stays isotropic throughout its evolution even if a central IMBH is present.
\begin{table*}
\begin{minipage}[t!]{\textwidth}
\caption{Initial parameters and the $\chi^2$ values for best-fit models containing different IMBH mass fractions.}
\label{table:1}     
\centering                          
\renewcommand{\footnoterule}{}  
\begin{tabular}{c c c c c c c c}        
\hline\hline                 
Model & $c$ & $r_{hp}$ (pc) & absolute $\chi^2_{rv}$ & absolute $\chi^2_{pm}$ & absolute $\chi^2_{sb}$ & reduced $\chi^2_{total}$\footnote{This column is obtained by dividing the sum of absolute $\chi^2$ on the total number of data points for all three observed profiles} & reduced $\chi^2_{total} ( <40^{\prime\prime})$\footnote{The same as the previous column but for the inner $40^{\prime\prime}$ region}\\    
\hline                      
   no-IMBH & 0.8 & 12.8 & 135.61 & 294.80 & 537.31 & 3.36 & 3.06\\   
   1\% IMBH & 0.7 & 12.8 & 99.91 & 345.52 & 313.21 & 2.63 & 3.80\\
{\bf 2\% IMBH} & {\bf 0.5} & {\bf 13.4} & {\bf 62.41} & {\bf 271.51} & {\bf 277.95} & {\bf 3.09} & {\bf 2.58}\\
   3\% IMBH & 0.4 & 13.4 & 58.93 & 325.88 & 654.72 & 3.61 & 5.01\\
\hline                                  
\end{tabular}
\end{minipage}
\end{table*}
\begin{figure}[] 
   \centering
   \includegraphics[width=0.5\textwidth]{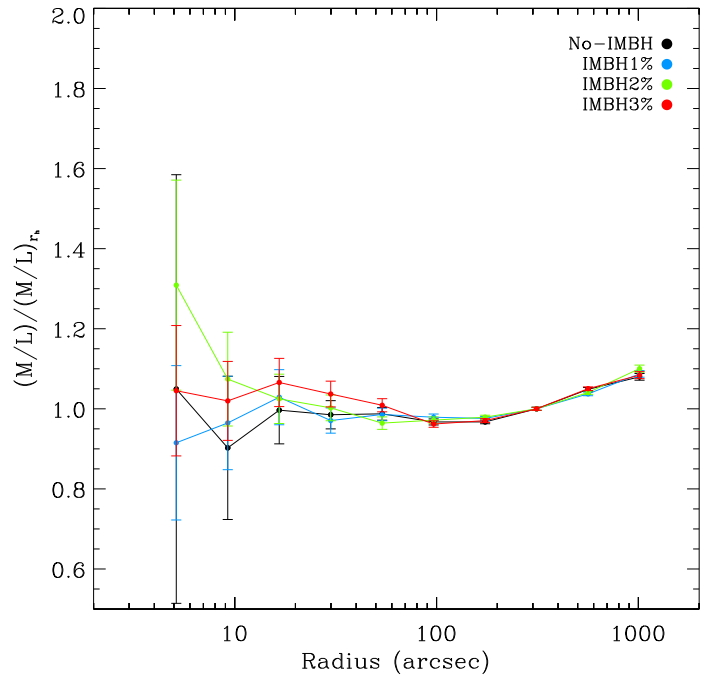}
      \caption{V-band stellar M/L ratio as a function of distance to cluster center for the best-fit models normalized by M/L at half-light radius. The M/L ratio increases slightly outward from the core radius due to an increase in the fraction of low mass stars.}
         \label{ML}
   \end{figure}

In Fig.~\ref{ML}, we show the stellar V-band mass-to-light ratio profiles in our simulations for the best-fit models with an IMBH, and the model without an IMBH. Since the absolute value of M/L depends on the age of the stars, we calculated it using the same 20 snapshots, i.e. from 11 to 12 Gyr, for all the models in Fig.~\ref{ML}. The M/L ratio is the same for all models at large radii since the initial mass function is the same, and, the effect of an IMBH is negligible in the outer region.  The M/L ratio seems to be constant in the central region and is increasing outward starting from the core radius ($\sim100^{\prime\prime}$). It is about 10\% larger than its value at the core radius for all models due to the increase in number of low mass stars at larger radii.

As mentioned in Section 3, setting the relaxation time and fixing the final present-day $r_{hp}$ to the one of $\omega$ Centauri, we obtain the final (present-day) total mass of the model cluster. We derive a cluster mass of $(2.6\pm0.1)\times10^6 M_{\odot}$ if we use the best-fit model containing an IMBH of 2\% of the cluster mass. This is in very good agreement with the mass of $\omega$ Centauri as
determined by vdV06, $(2.5 \pm 0.3)\times10^6 M_{\odot}$.
\section{Discussion and Conclusions}
We have created a large set of evolutionary N-body models for $\omega$ Centauri assuming that the cluster started with a Kroupa IMF, with the aim of reproducing its observed properties. The main goal is to check models using standard assumptions as a first attempt, and to see to what extent we reproduce the observations. In particular, we examine whether we can explain the newly acquired observations for the central velocity dispersion profile with the presence of a central IMBH or whether a model without an IMBH is also consistent with the observations. 

Following the method applied in Baumgardt et al. (2003a,b), we calculate models starting with spherical isotropic King conditions (King 1966) with different initial parameters. We do not include the tidal field of the Galaxy and primordial binaries in our models. Since we cannot simulate a star cluster of the size of $\omega$ Centauri by direct N-body simulations, we start with more extended clusters containing fewer number of stars than $\omega$ Centauri and scale our model clusters to the observed cluster such that the relaxation time is constant. As described in detail in Section 3, we measure physical quantities from our models such as velocity dispersion and surface brightness following the methods in observational studies, with the same magnitude cut-offs and luminosity weights. Using such careful magnitude weights and radial cut-offs in our models make the data-model comparisons and consequently the drawn conclusions more reliable. We use $\chi^2$ values to compare the profiles of model clusters after 12 Gyr of evolution with the observed ones. In Section 4, we present a grid of models for clusters containing an IMBH mass of 2\% of the cluster stellar mass.

We show that the best-fit IMBH model, containing a $5\times10^4 M_{\odot}$ black hole, matches the data presented in Noyola et al. (2010) very well. In particular, we reproduce the observed rise in the central velocity dispersion as an indicator for the presence of an IMBH. We stress that relying on the profiles relative to the kinematic center in Noyola et al. (2010) makes it impossible to consistently fit the radial velocity dispersion over all radii without a central IMBH. Furthermore, we show that M/L is increasing from the core radius towards large radii for all models independent of the presence of a central black hole. We do not claim that our best-fit model is unique. However, we examine more than 100 models to be confident about the initial parameters and the final chosen best-fit model.

A number of further details can be investigated or improved as a next step. Among the interesting issues are the study of rotation at different radii, especially in the central region. In principle, such a study can be done using an axisymmetric model which is closer to $\omega$ Centauri, as an exception in Galactic star clusters.
 \citet{2010MNRAS.405..194F} investigated the evolution of rotating dense stellar systems containing massive black holes. Exploring rotation effects could help to better understand the observed discrepancy between proper motion and radial velocity dispersion. In addition, as described in previous studies \citep{2002ASPC..265...21K, 1987A&A...184..144M}, $\omega$ Centauri might not be well fitted by a King profile alone. We intend to investigate this in more detail in a future paper studying different initial configurations such as Sersic and double King profiles. Furthermore, to improve the model surface brightness profile at large radii, around the tidal radius, the tidal field of the Galaxy should be taken into account. In order to more tightly constrain the initial parameters for $\omega$ Centauri the above studies as well as trying different IMF distributions are necessary in addition to investigate a finer grid, though they will be computationally expensive.
\begin{acknowledgements}
This work was supported by the DFG cluster of excellence ‘Origin and Structure of the Universe’.
H.B. acknowledges support from the German Science foundation through a Heisenberg Fellowship and from the
Australian Research Council through Future Fellowship grant FT0991052. B.J. and H.B. would like to thank Roeland van der Marel and Jay Anderson for providing their data prior to publication. K.G. acknowledges support from NSF-0908639.
\end{acknowledgements}

\bibliographystyle{bibtex/aa}
\bibliography{omegacen}

\end{document}